\documentclass[letterpaper,usenatbib]{mn2e}
\newcommand{\ergs}{\ifmmode {\rm erg\ s}^{-1} \else erg s$^{-1}$\ \fi}

\newcommand{\feii}{Fe {\sc ii}\ }
\newcommand{\mgii}{Mg {\sc ii}}
\newcommand{\civ}{C {\sc iv}\ }
\newcommand{\ciii}{C {\sc iii]}\ }

\newcommand{\siiv}{Si {\sc iv}\ }

\newcommand{\lb}{\ifmmode L_{\rm Bol} \else $L_{\rm Bol}$\ \fi}
\newcommand{\ledd}{\ifmmode L_{\rm Edd} \else $L_{\rm Edd}$\ \fi}
\newcommand{\leddR}{\ifmmode L_{\rm Bol}/L_{\rm Edd} \else $L_{\rm Bol}/L_{\rm Edd}$\ \fi}
\newcommand{\lx}{\ifmmode L_{\rm 2-10keV} \else  $L_{\rm 2-10keV}$\ \fi}
\newcommand{\hb}{\ifmmode H\beta \else H$\beta$\ \fi}
\newcommand{\ha}{\ifmmode H\alpha \else H$\alpha$\ \fi}
\newcommand{\hg}{\ifmmode H\alpha \else H$\gamma$\ \fi}

\newcommand{\heii}{He {\sc ii}\ }
\newcommand{\mbh}{\ifmmode M_{\rm BH}  \else $M_{\rm BH}$\ \fi}
\newcommand{\lv}{\ifmmode \lambda L_{\lambda}(1350\AA) \else $\lambda L_{\lambda}(1350\AA)$\ \fi}
\newcommand{\lcon}{\ifmmode L_{1350} \else $L_{1350}$\ \fi}

\newcommand{\mdot}{\ifmmode \dot{m} \else \dot{m} \fi }
\newcommand{\llog}{\ifmmode {\rm log} \else {\rm log} \fi }
\newcommand{\kms}{\ifmmode {\rm km\ s}^{-1} \else km s$^{-1}$\ \fi}

\usepackage{graphicx}
\usepackage{color}
\usepackage{lscape}
\usepackage{hyperref}

\begin{document}
\title[BAL variability]{Variability of broad absorption lines in a
QSO SDSS J022844.09+000217.0 on multi-year timescales}
\author[He, Bian, Jiang \& Wang]{Zhi-Cheng He,
Wei-Hao Bian\thanks{E-mail: whbian@njnu.edu.cn},
Xiao-Lei Jiang \& Yue-Feng Wang \\
Department of Physics and Institute of Theoretical Physics, Nanjing
Normal University, Nanjing 210023, China\\} \maketitle

\begin{abstract}
The variability of the broad absorption lines is investigated for a
broad absorption line (BAL) QSO, SDSS J022844.09+000217.0 (z =
2.719), with 18 SDSS/BOSS spectra covering 4128 days in the
observed frame. With the ratio of the rms spectrum to the mean
spectrum, the relative flux change of the BAL-trough is larger than
that of the emission lines and the continuum. Fitting the power-law
continuum and the emission line profiles of \civ $\lambda$1549 and
\siiv$\lambda$1399, we calculate the equivalent width (EW) for
different epochs, as well as the continuum luminosity and the spectral
index. It is found that there is a strong correlation between the
BAL-trough EW and the spectral index, and a weak negative
correlation between the BAL-trough EW and the continuum luminosity. The
strong correlation between the BAL-trough EW and the spectral index
for this one QSO suggests that dust is intrinsic to outflows. The weak
correlation between the BAL variability and the continuum luminosity
for this one QSO implies that the BAL-trough variation is not dominated
by photoionization.
\end{abstract}


\section{INTRODUCTION}
Broad absorption line quasars (BAL QSOs) exhibit broad, ultraviolet
(UV) line absorption troughs. The traditional BAL QSOs were
quantified by balnicity index (BI) with velocity width larger than
2000 \kms and outflow velocity of 3000-25000 \kms \citep{Weymann91}. 
There are a large number of broad absorption features within
3000 \kms and beyond 25000 \kms, as well as features with widths
less than 2000 \kms, and newer ways of quantifying the index 
\citep[e.g.][]{Hall02, Trump06, Gibson09}. BAL troughs
are present in about 10-40\% of QSOs \citep[e.g.][]{Hewett03, Trump06, Ganguly07, Allen11},
produced by absorption from high-ionization lines such as \siiv
1399, \civ 1549, \ciii 1909 (known as HiBAL), and low-ionization
lines such as \mgii 2799 (known as LoBAL). The fraction of BAL QSOs
strongly depends on the definition of BAL QSOs \citep[e.g.][]{Hewett03, Trump06}, 
so drawing physical conclusions from
population statistics must be done with care.

BAL troughs are thought to be the strongest observed signatures of
QSO winds \citep{Fabian12}. The observation of BALs is commonly thought
to be the result of passing a disk wind along the line of sight
\citep[e.g.][]{Murray95, Elvis00}. If the wind is on the line
of sight, a QSO appears as a BAL QSO. That detection of a wind
should be orientation-dependent is very similar to the case of other
structures in QSOs, such as the broad line region (BLR) or jet
\citep[e.g.][]{Urry95}.

However, it was found that there is no significant difference
between radio spectral index distributions for radio-loud BALs and
radio-loud non-BALs, indicating they have similar ranges of viewing
angles \citep{Becker00, Mon08, Fine11, Bruni12}. No correlations exist between outflow
properties and the orientation, suggesting that BAL winds along any
line of sight are driven by the same mechanisms (DiPompeo et al.
2012). There are also BAL QSOs that are seen very nearly along the
radio jet axis \citep{Zhou06, Ghosh07}.
Another explanation for BALs is an orientation-independent evolution
effect \citep[e.g.][]{Gibson08, Zubovas13}. BALs outflow
is possibly caused by the expulsion of gas and dust by galaxy
collision as an evolutionary stage of AGN \citep[e.g.][]{Voit93,
Gregg06, Gibson08}.

The disk wind is believed to come from the central accretion disk in
QSOs \citep[e.g.][]{Murray95}. The dependence of wind properties
(such as the trough EW, the maximum velocity, the fraction of BAL
QSOs) on the physical properties of QSOs (such as the luminosity,
the \heii $\lambda 1640$ \AA\ EW, and the UV continuum slope) has been
investigated by many authors, and they found some correlations
between them \citep{Laor02, Ganguly07, Baskin13}. It was found that the maximum outflow velocities increase
with both the bolometric luminosity and the blueness of the spectral
slope, suggesting the idea of radiation pressure-driven outflows
\citep{Laor02, Ganguly07}.

BAL troughs often vary in equivalent width (EW) and/or shape over
rest-frame timescales of months to years \citep[e.g.][]{Barlow92,
Gibson08, Filiz12, Filiz13}. It was found that
the fractional EW change increases with rest-frame timescale over
the range 0.05 - 5 yr \citep[e.g.][]{Gibson08, Filiz13}. The BAL-trough variation is believed to be driven by changes
in velocity structure,  covering factor, or ionization level
\citep[e.g.][]{Filiz12}. It was suggested that BAL variation arises from
changes in the amount of "shielding gas" along the line-of-sight
\citep{Filiz13}.. Such changes are likely according to disk-wind
simulations \citep[e.g.][]{Proga00, Murray95}. These changes
could ultimately be related to the central accretion disk \citep[e.g.][]{Proga00}. 
Changes in the column density of shielding gas
(BAL trough EW variability) can relate to the level of ionizing
luminosity reaching the BAL wind.

Two kinds of investigations of BAL-trough variability have typically
been undertaken. One
is to use two-epoch spectra in a BAL sample; the other is to use
multi-epoch spectra for one single object \citep[e.g.][]{Gibson08,
Filiz13, Capellupo11, Capellupo12, Capellupo13}. Using a
two-epoch sample of 13 BAL QSOs overlapping between the Very Large Array
(VLA) Faint Images of the Radio Sky at Twenty-Centimeter (FIRST) Survey and
the Sloan Digital Sky Survey (SDSS), \cite{Gibson08} found that BALs
tend to vary on multi-year timescales in velocity regions that are a
few thousand kilometers per second wide. They also found that
BAL-trough variations do not appear to be correlated with variations
in the observable continuum, implying that the BAL-trough variation
is not dominated by photoionization changes \citep[e.g.][]{Barlow89,
Capellupo12}. \cite{Capellupo11, Capellupo12, Capellupo13} did
extensive variability studies on a small BAL QSO sample with longer
timescales and multiple epochs. They found that the variability occurs
typically in only portions of the BAL troughs; the components at
higher outflow velocities are more likely to vary than those at
lower velocities, and weaker BALs are more likely to vary than
stronger BALs. The shortest timescales constrain the
location of the outflowing gas. With a larger SDSS sample, \cite{Filiz13} found that there is a relation between the
luminosity and EW variability on moderate timescales, although they
did not regard this case as strong evidence for luminosity
dependence.

The BAL-trough variability for one single object is very important
in understanding of the origin of the BAL QSOs. With the SDSS, a
large number of BAL QSOs are found in various data releases 
\citep[e.g.][]{Trump06, Gibson09, Shen11, Paris13}. 
We selected a single QSO, SDSS J022844.09+000217.0, which
has the highest number of observations in the time domain, to
investigate the relation between the wind and QSO properties. \S 2
presents the data. \S 3 gives the data analysis. \S 4 contains our
results and discussion. The Summery is given in the last section.
Throughout this work we use a cosmology with $H_0 = 70 \kms \rm
Mpc^{-1}$, $\Omega_M = 0.3$, and $\Omega_{\Lambda} = 0.7$.

\section{SDSS/BOSS spectral Data}
The Baryon Oscillation Spectroscopic Survey (BOSS), part
of SDSS-III, uses the dedicated 2.5-m wide-field telescope at Apache
Point Observatory near Sacramento Peak in Southern New Mexico to
conduct an imaging and spectroscopic survey for about 1.5 million
luminous galaxies as well as about 160,000 quasars at $z > 2.2$
\citep{Eisenstein11}. With respect to the original SDSS, the
wavelength coverage of BOSS spectra changes from 3800 \AA\ -9200
\AA\ to 3600 \AA\ -10400 \AA, as well as the fiber numbers per plate
from 640 to 1000. And the BOSS uses new fibers (1000 rather than 640
per plate), with smaller holes (2" rather than 3").

With SDSS DR7, \cite{Shen11} gave a compilation of properties
of the 105783 QSOs. There are 6214 BAL QSOs in their SDSS DR7 sample
\citep[see also][]{Gibson09}. There are 456 \civ BAL QSOs with at
least two-epoch spectroscopic observations. We searched for observations of
these 456 \civ BAL QSOs in SDSS DR10 and found
more epochs of spectroscopy for each object. The one BAL QSO with the most
observations, SDSS J022844.09+000217.0 (z = 2.719), has 18 high
signal-to-noise ratio (S/N) SDSS/BOSS spectra, covering 4128 days
(about 11 years) in the observed frame (Table 1), i.e., about 3
years in the rest frame. For SDSS spectra, the S/N in the r-band is
about 17-20. The BOSS spectra have longer exposure times than the SDSS
sample, and have a better S/N about 22-36. In Table 1, the
information for these 18 spectra is listed, such as the survey name,
MJD, Plate, Fiber ID, S/N.

\begin{table*}
\centering \caption{The properties of 18-epoch spectra for SDSS
J022844.09+000217.0. Col(6-9): in units of $\AA$. Col(10): measured from  1300\AA\ to 2400\AA ,in units
of $10^{43} \ergs \AA^{-1}$. Col(12): in units of $\kms$ . $EW_{1}$
is measured based on a power-law continuum. $EW_{2}$ is measured
based on a pseudo-continuum of power-law continuum plus
emission lines.}
\begin{tabular}{lccccccccccccccccccc}
\hline \hline
Survey & MJD & Plate & Fiber &  SN(r)& $EW_{1}$(\civ) & $EW_{2}$(\civ) & $EW_{1}$(\siiv) & $EW_{2}$(\siiv)& $L_{cont}$ & $\alpha$ & $V_{max}$ \\
(1)&(2)&(3)&(4)&(5)& (6) & (7) & (8) & (9) & (10) & (11) & (12) \\
\hline
SDSS & 51817 & 406 & 35 & 17.1 & $12.86\pm 0.79$ & $18.14\pm 1.23$ & $8.59\pm 0.55$ & $9.48\pm 0.62$ & $3.69\pm 0.16$ & $-0.76\pm 0.02$ & 12738\\
SDSS & 51869 & 406 & 37 & 19.6 & $15.33\pm 0.91$ & $19.32\pm 1.12$ & $9.77\pm 0.49$ & $10.66\pm 0.54$&  $3.20\pm 0.13$ & $-1.16\pm 0.03$ & 13124 \\
SDSS & 51876 & 406 & 37 & 17.0 & $14.35\pm 1.01$ & $18.13\pm 1.24$ & $8.90\pm 0.58$ & $10.26\pm 0.67$&  $3.50\pm 0.15$ & $-0.79\pm 0.03$  & 12738 \\
SDSS & 51900 & 406 & 36 & 17.0 & $14.96\pm 0.80$ & $19.58\pm 1.32$ & $9.29\pm 0.57$ & $10.08\pm 0.63$&  $3.72\pm 0.17$ & $-0.74\pm 0.03$ & 12738\\
SDSS & 52200 & 705 & 431 & 15.6& $15.62\pm 1.21$ & $20.07\pm 1.52$ & $9.23\pm 0.63$ & $10.29\pm 0.70$&  $2.99\pm 0.16$ & $-0.75\pm 0.03$ & 13317\\
SDSS & 52205 & 704 & 640 & 16.6& $16.00\pm 1.03$ & $19.57\pm 1.35$ & $8.93\pm 0.52$ & $9.64\pm 0.57$ &  $2.94\pm 0.16$ & $-0.90\pm 0.03$ & 13124\\
SDSS & 52238 & 406 &  39 & 19.3& $16.70\pm 1.11$ & $20.55\pm 1.35$ & $9.20\pm 0.62$ & $9.82\pm 0.67$ &  $3.59\pm 0.13$ & $-0.77\pm 0.03$ & 13703\\
\hline
BOSS & 55179 & 3615 & 780 & 36.6 & $17.02\pm 0.54$ & $22.67\pm 0.71$ & $10.97\pm 0.31$ & $12.32\pm 0.35$ &  $3.03\pm 0.07$ & $-0.32\pm 0.02$ & 13703\\
BOSS & 55181 & 3647 & 778 & 28.6 & $16.10\pm 0.67$ & $22.33\pm 0.91$ & $10.16\pm 0.38$ & $11.33\pm 0.43$ &  $3.16\pm 0.09$ & $-0.32\pm 0.02$ & 13703\\
BOSS & 55208 & 3615 & 780 & 29.0 & $16.24\pm 0.59$ & $21.60\pm 0.77$ & $10.18\pm 0.34$ & $11.16\pm 0.38$&  $2.67\pm 0.07$ & $-0.67\pm 0.02$ & 13703 \\
BOSS & 55209 & 3744 & 575 & 22.3 & $15.15\pm 0.83$ & $20.35\pm 1.10$ & $10.08\pm 0.51$ & $11.03\pm 0.49$ &  $2.90\pm 0.09$ & $-0.62\pm 0.03$ & 13896\\
BOSS & 55241 & 3647 & 778 & 33.8 & $16.46\pm 0.57$ & $22.56\pm 0.77$ & $10.33\pm 0.33$ & $11.47\pm 0.35$ &  $3.19\pm 0.07$ & $-0.46\pm 0.02$ & 14089 \\
BOSS & 55445 & 3615 & 770 & 27.7 & $15.94\pm 0.73$ & $21.61\pm 0.97$ & $9.97\pm 0.41$ & $11.17\pm 0.36$  &  $2.87\pm 0.09$ & $-0.41\pm 0.02$ & 14089 \\
BOSS & 55455 & 4238 & 736 & 30.8 & $16.40\pm 0.63$ & $22.09\pm 0.83$ & $9.56\pm 0.32$ & $10.50\pm 0.42$ &  $3.13\pm 0.09$ & $-0.73\pm 0.02$  & 14089\\
BOSS & 55476 & 3647 & 786 & 34.0 & $16.58\pm 0.59$ & $22.54\pm 0.78$ & $10.34\pm 0.32$ & $11.49\pm 0.40$ &  $3.43\pm 0.09$ & $-0.54\pm 0.02$  & 13896\\
BOSS & 55827 & 3647 & 780 & 37.0 & $17.69\pm 0.67$ & $22.80\pm 0.86$ & $10.86\pm 0.38$ & $12.07\pm 0.43$ &  $3.22\pm 0.10$ & $-0.62\pm 0.02$ & 14089\\
BOSS & 55856 & 3615 & 776 & 32.5 & $17.83\pm 0.80$ & $22.94\pm 1.04$ & $11.27\pm 0.44$ & $12.74\pm 0.50$ &  $3.16\pm 0.12$ & $-0.40\pm 0.02$ & 14089 \\
BOSS & 55945 & 3647 & 738 & 35.4 & $16.11\pm 0.65$ & $22.14\pm 0.90$ & $10.67\pm 0.40$ & $11.95\pm 0.45$ &  $3.53\pm 0.12$ & $-0.44\pm 0.02$  & 13896\\
\hline
\end{tabular}
\end{table*}

\begin{table*}
\centering \caption{Summary of the Spearman correlation
coefficients: $EW_{1}$ is measured based on a power-law continuum.
$EW_{2}$ is measured based on a pseudo-continuum of a power-law
continuum plus emission lines. $EW_{h}$ is measured within the
high-velocity part. $EW_{l}$ is measured within the low-velocity
part. The first two lines are for the \civ BAL-trough, and the last two
lines are for the \siiv BAL-trough. The value in brackets is the
probability of the null hypothesis. $L_{cont}$ is measured from
1300\AA\ to 2400\AA.}
\begin{tabular}{lccccccccccccccccccc}
\hline \hline
 & & $EW_{1}$ & $EW_{2}$ &$EW_{h}$ & $EW_{l}$\\
\hline
 \civ & $L_{cont}$ & -0.19 (0.443) & -0.21 (0.399)   & -0.22 (0.39)   & -0.28 (0.26)\\
      & $\alpha$   &  0.45 (0.06)  &  0.77 (0.00018) &  0.64 (0.0042) & 0.87 (0.000003) \\
\hline
\siiv & $L_{cont}$ &  -0.24 (0.33)     &-0.22 (0.37)   & -0.35 (0.15)    & -0.22 (0.39)   \\
      & $\alpha$   &   0.78 (0.00013)  & 0.81 (0.00004)&  0.75 (0.0003)   & 0.82 (0.00003)  \\
\hline
\end{tabular}
\end{table*}

\section{Analysis}
\subsection{Mean and RMS spectra}

\begin{figure}
\begin{center}
\includegraphics[height=8.0cm,angle=-90]{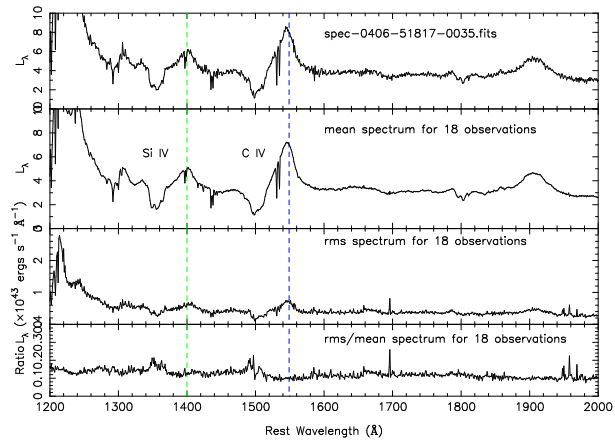}
\caption{A typical single spectrum (top panel), the average spectrum
(second panel), the root-mean-square (rms) spectrum (third panel),
the ratio of the rms spectrum to the average spectrum (i.e., the
relative flux change, bottom panel) from total 18 spectra for SDSS
J022844.09+000217.0. The dotted lines show the wavelengthes of \siiv
$\lambda$ 1400 \AA\  and \civ $\lambda$ 1549 \AA.}
\end{center}
\end{figure}

The mean spectrum from these 18 spectra is calculated as follows:
$$f_{mean}(\lambda )=\sum_{i=1}^{N} \frac {f_{i}(\lambda )}{N}$$
where the sum is taken over the N = 18 spectra. From the mean
spectrum with high S/N, the absorption region and emission region
are more obvious, which can help us choose suitable continuum
windows. The root-mean-square (rms) spectrum from the 18 spectra is
determined by:
$$f_{rms}(\lambda )= \left\{ \frac{1}{N-1} \sum_{i=1}^{N} [f_{i}(\lambda)
-f_{mean}(\lambda )]^{2} \right\}  ^{1/2}$$ where the sum is taken
over the N = 18 spectra (e.g., Kaspi et al. 1999). For the spectrum
at MJD = 52205, the data from 1660 \AA\ -- 1798 \AA\ (rest
wavelength) are not used. The mean spectrum and the rms spectrum are
shown in Fig. 1., as well as a single spectrum (top panel).

From the mean spectrum (second panel in Fig. 1), the broad
absorption troughs in \siiv and \civ are obvious, as well as some
narrow absorption lines. From the rms spectrum, the emission region
varied significantly larger than the absorption region (third panel
in Fig. 1). The rms-to-mean ratio spectrum, i.e., the relative flux
change, is shown in the bottom panel of Fig. 1. The variance
fraction from these 18 spectra increases as the frequency increases,
from about 10\% on the red side to about 15\% on the blue side. The
relative flux change is larger for the absorption trough than for
the continuum and the emission lines.

In Fig. 2, we also show all 18 spectra normalized at 2000 \AA.
There is clear variability of the spectral slope during this long
timescale of about 11 years.

\begin{figure*}
\begin{center}
\includegraphics[height=8.0cm,angle=-90]{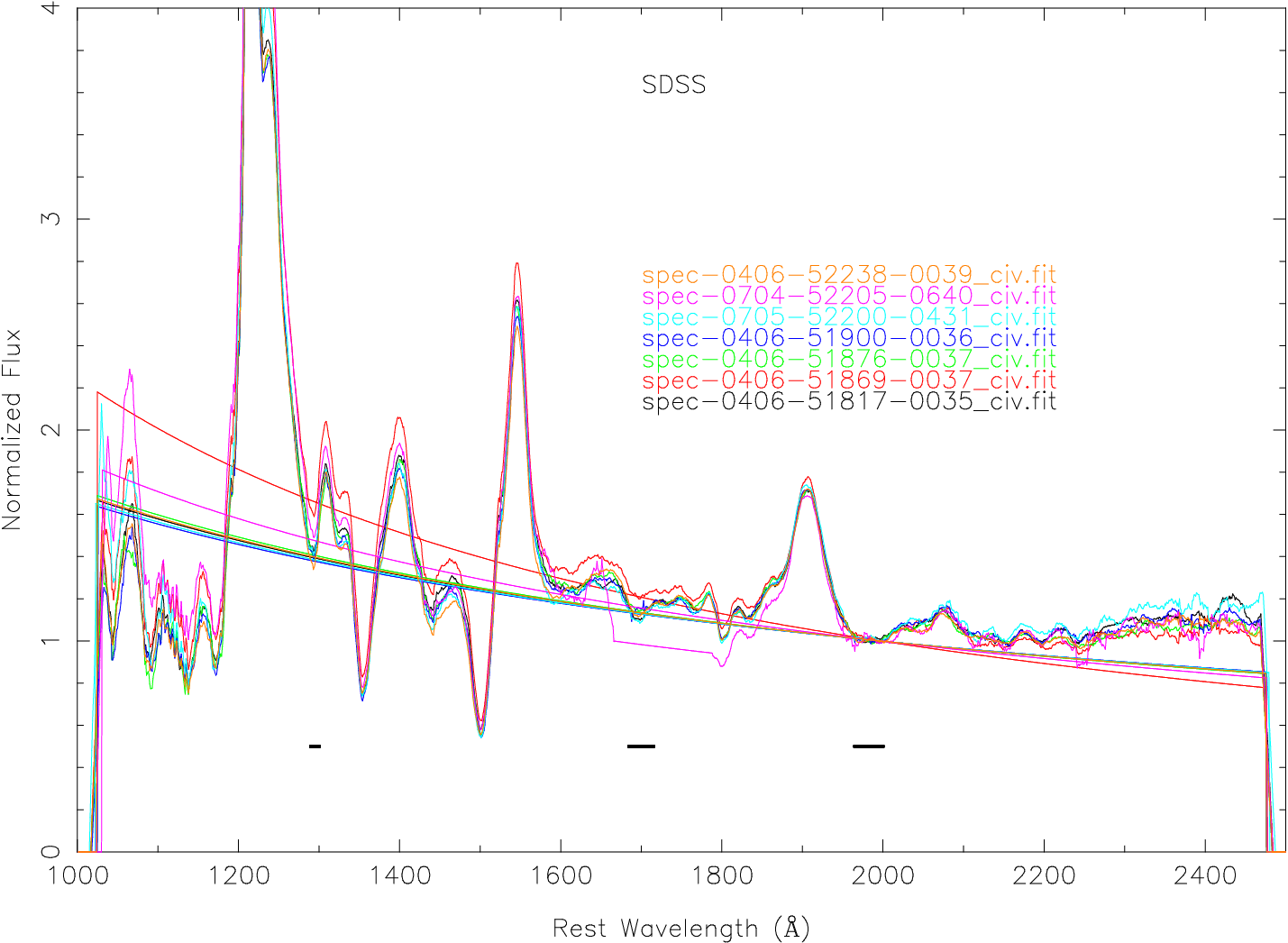}
\includegraphics[height=8.0cm,angle=-90]{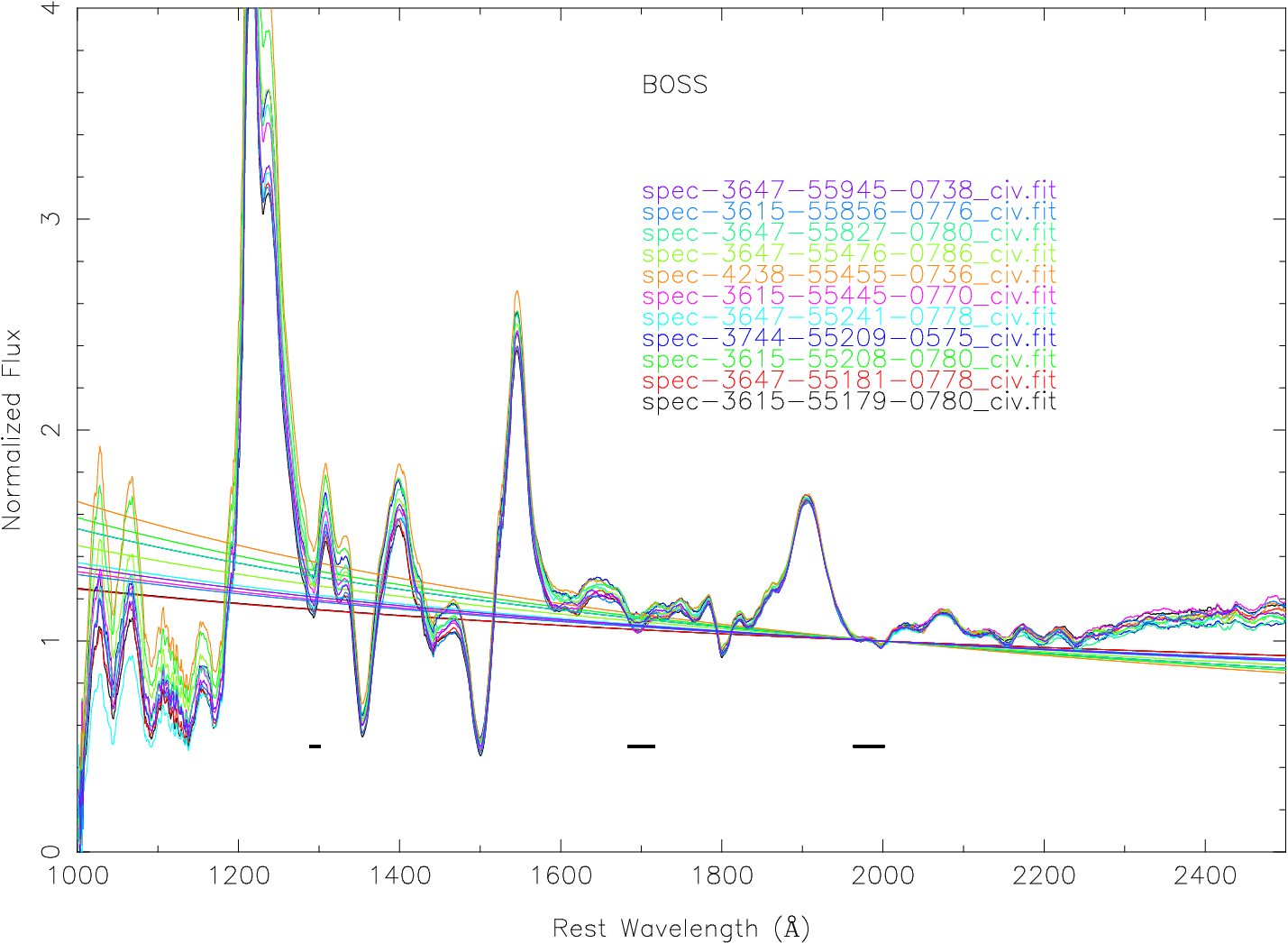}
\caption{All 7 SDSS (left) and 11 BOSS (right) spectra normalized
at 2000 \AA\ with a smoothing box of 20 \AA. The power-law continuum is
plotted in this figure. The black horizontal lines are the continuum
windows for the power-law fit.}
\end{center}
\end{figure*}

\subsection{Fitting the spectrum}
\begin{figure*}
\begin{center}
\includegraphics[width=6cm,angle=-90]{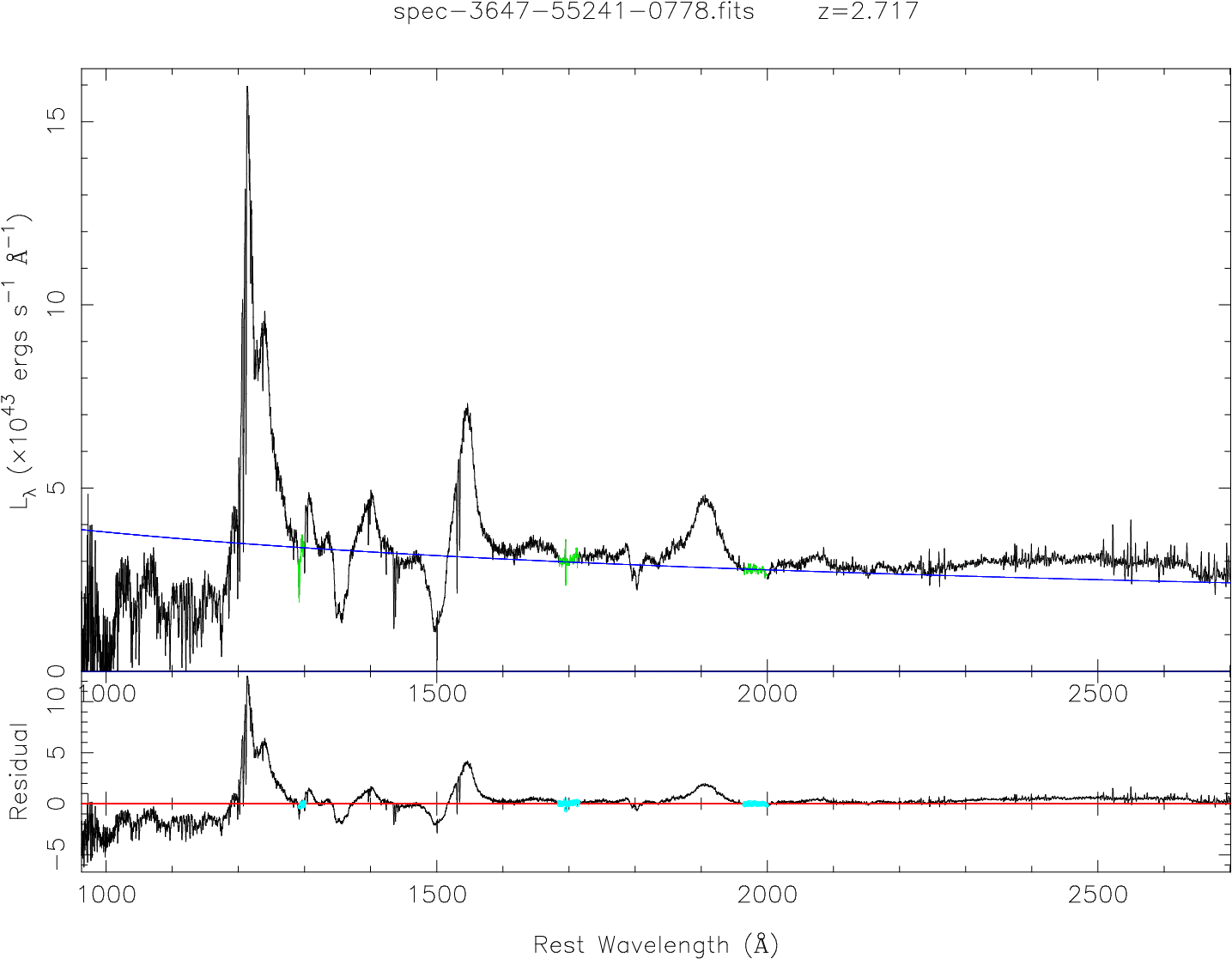}
\includegraphics[width=6cm,angle=-90]{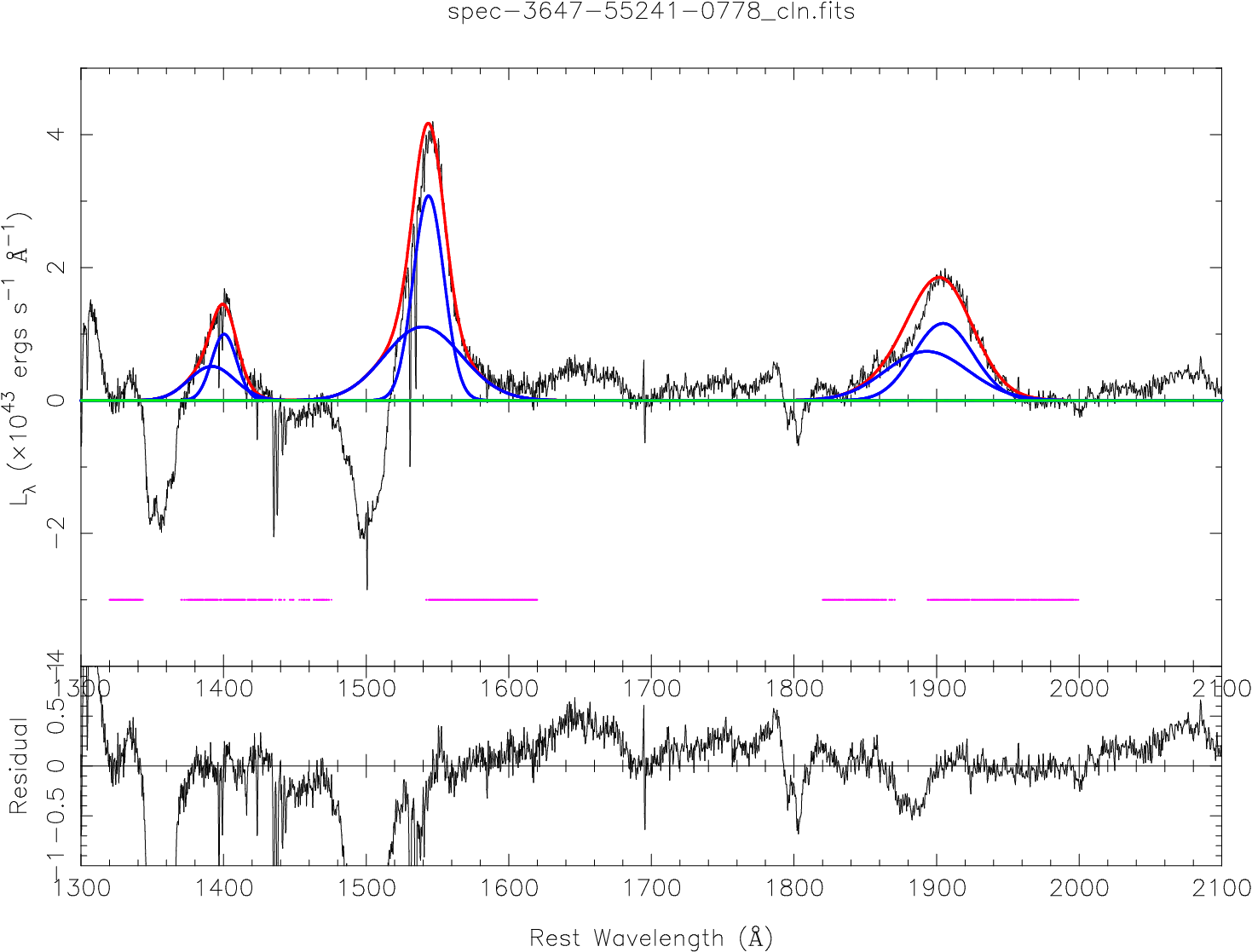}
\caption{ An example of the power-law continuum fit (left) and the emission
line fit (right). Left top: The extinction-corrected rest-frame
spectrum is shown by the black lines. The green dots are the initial
continuum windows. The blue line is the power-law continuum. Left
bottom: the residual spectrum. Right top: The multiple Gaussian
components are in blue and the sum of these is in red. The fitting
window is in blue. The horizontal pink dots are the final data used in
emission line fit. Right bottom: the residual spectrum. }
\end{center}
\end{figure*}

In order to investigate the relation between the BAL-trough and QSO
properties, we fit the continuum and the emission line profiles as
follows \citep{Hu08, Zhang10, Bian12, Capellupo12}:

(1) Fitting the continuum. All the observed spectra are corrected
for Galactic extinction using an $A_V$ value of 0.1575, assuming the
extinction curve of Cardelli et al. (1989; IR band; UV band) and
O'Donnell (1994; optical band) with $R_V = 3.1$. And they are
converted to the rest frame by redshift of z = 2.717. It is common
to use the power-law formula, $f_\lambda \propto \lambda ^{\alpha} $
($f_\upsilon \propto \upsilon ^{-(2+\alpha)} $) to approximately fit
the QSO continuum spectrum. Other continuum models, such as a
polynomial function or a reddened power-law, have also been used in
previous studies \citep{Lundgren07, Gibson09}. We fit
the power-law continuum iteratively in the "continuum windows",
which are known to be relatively free from strong emission lines.
The fit is performed by minimizing $\chi^2$. From Fig. 1, we select
the following continuum windows: 1290-1300, 1685-1715, 1965-2000
\AA\ in the rest frame \citep[e.g.][]{Forster01, Vanden01, Gibson08, Bian12, Baskin13}.
At each iteration, we ignore any spectral bins that deviate by more
than 3$\sigma$ from the previous iteration's continuum fit. That
iterative approach would exclude additional spectral regions that
contain broad emission or absorption features, especially for BAL
QSOs. For this object, we don't fit the UV \feii emission. The
power-law continuum is also shown in Fig. 3. An example of the
continuum fit and the residual is shown in the left panel of Fig. 3.

(2) Fitting the broad emission lines. Considering the broad wing in
the broad emission line profiles, we use three sets of two Gaussians
to simultaneously model emission line profiles for \siiv $\lambda$
1399, \civ $\lambda$ 1549, \ciii $\lambda$ 1909 in the
continuum-subtracted spectra. Each emission line is fit iteratively.
At each iteration, the "absorbed" bins are ignored when they are
more than 2.5$\sigma$ below the previous model fit. In order to
eliminate absorption trough impact on the fitting of the emission
lines of the \civ, \siiv, \ciii, especially in the blue wings of
these lines, the fitted weights of the blue wings for these emission
lines are set to half of normal. And the weights of the red wings
are set to twice normal. An example of the line fit and the
residual is shown in the right panel of Fig. 3. The horizontal pink dots
are for the line fitting. The red lines show the sum of the blue Gaussian
profiles for the different lines.

\subsection{The BAL EW and the maximum velocity of outflow}
Two methods are used to calculate the BAL-trough EW. One is based on
the power-law continuum, the other is based on the pseudo-continuum
of the power-law plus the emission lines. The latter includes the
BAL correction from the emission lines. The equivalent widths of the
BAL-troughs for \siiv and \civ are calculated as follows:
$$EW =\int_{}^{} [ 1-\frac {f_{obv}(\lambda)}{f_{con}(\lambda)} ] d\lambda $$
The integration is done for $f_{obv}(\lambda) < f_{con}(\lambda)$.
From Fig. 5, for the \siiv  $\lambda$ 1399 BAL, its EW is integrated
from 1340 \AA\ to 1400 \AA, and for the \civ $\lambda$ 1549 BAL, its
EW is integrated from 1450 \AA\ to 1550 \AA. The EW results are
shown in Cols. 6-9 in Table 1, where the subscript of 1 is for the
case based on the power-law continuum ($EW_{1}$), and the subscript
of 2 is for the case based on the pseudo-continuum of the power-law
plus the emission lines ($EW_{2}$). The latter EW is larger than the
former one. The $EW_{2}$ are about 10\% larger than the $EW_{1}$ for
\siiv, while the difference is about 25\% for \civ.

We also measure the maximum velocity of outflow. We searched each
spectrum for absorption troughs at a level of 10\% below the
continuum as given in the traditional definition of a BAL. The maximum
velocity corresponds to the shortest wavelength for a given trough
with at least 10 \AA\ deeper than 10\% below the continuum. The result is
listed in Col. (12) in Table 1.

For the power-law continuum, $f_{\lambda}=f_{2000}(
{\lambda/2000\AA})^{\alpha}$, the error for the power-law continuum
at the wavelength of $\lambda$ by error propagation is:
$$\delta(f_{con})=f_{\lambda}\sqrt{(\frac{\delta(f_{2000})}{f_{2000}})^2
+(\ln\lambda-\ln 2000)^2\delta\alpha^2}$$ where the errors of
$\delta(f_{2000}$) and $\delta\alpha$ are given in the power-law
fitting. The error for the BAL EW is measured as followed:
$$\delta(EW)=
\sqrt{\sum_{\lambda} (\frac{f_{obv}}{f_{con}})^2
[(\frac{\delta(f_{obv})}{f_{obs}})^2+(\frac{\delta(f_{con})}{f_{con}})^2]}$$
where $\delta(f_{obv})$ at $\lambda$ is the error in the SDSS/BOSS
spectrum.

\section{RESULTS AND DISCUSSION}

\subsection{Relation between the \civ emission line luminosity and the continuum}
There is a correlation between the \civ luminosity and the
continuum, where the continuum is the mean power-law luminosity
between 1300 \AA\ and 2200 \AA. The Spearman coefficient R is 0.67
with the the probability of the null hypothesis of $P_{null} =
0.002$. It is consistent with the photoionization scenario and
classic Baldwin effect for the emission lines in QSOs.

\subsection{The light curves for BAL-trough \civ EW}
\begin{figure}
\begin{center}
\includegraphics[height=8cm,angle=-90]{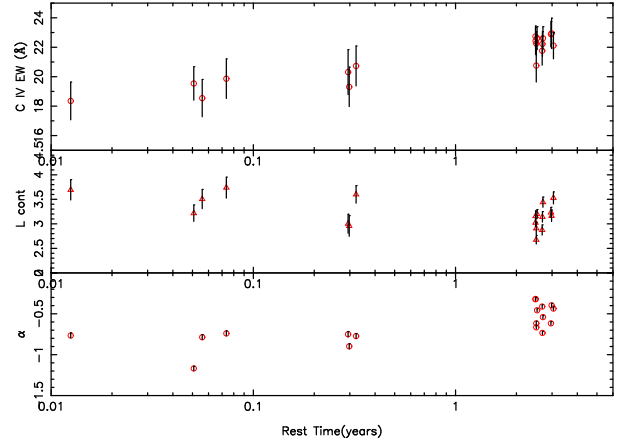}
\caption{The light curves for \civ BAL EW, the continuum (measured
from 1300 \AA\ to 2400 \AA) and $\alpha$. The continuum is in units
of $10^{43} \ergs \AA^{-1}$.}
\end{center}
\end{figure}

\begin{figure}
\begin{center}
\includegraphics[height=8cm,angle=-90]{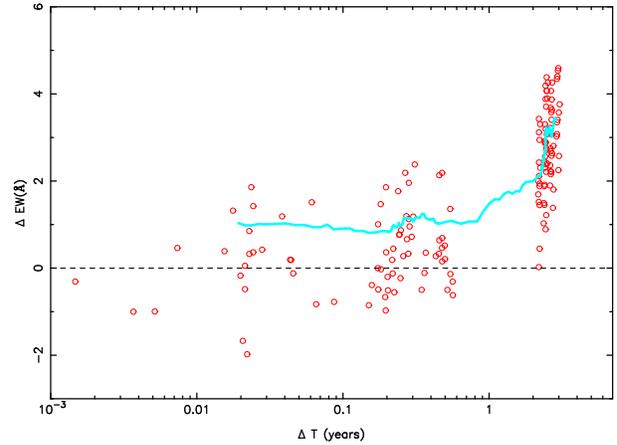}
\caption{\civ BAL EW variation versus timescale for two-epoch
spectra in the rest frame. The cyan line is the standard deviation
derived from the data, calculated using a sliding window containing
20 time-ordered data points. The standard deviation of EW variation
increases with increasing rest-frame timescale.}
\end{center}
\end{figure}

Fig. 4 gives the light curves of \civ EW, as well as the continuum,
and the spectral index $\alpha$ (from top to bottom panel). The
shortest time interval is one day, and the longest time interval is
about 11 years in the observed frame. With time, there is
an increasing trend of \civ EW and the spectral index, and a decreasing
trend of the continuum. The largest EW changes are about 25\% (from
18.1 \AA\ to 22.7 \AA) for \civ and 28\% (from 8.6 \AA\ to 11.0 \AA)
for \siiv.

It was found that \civ BAL-trough variability is larger for longer
timescales \citep[e.g.][]{Gibson08, Capellupo12, Filiz13}. We also give the EW variation versus the time
interval for two-epoch spectra (Figure 5). For longer time
intervals, the EW variation is larger for this object. That result
is consistent with previous studies \citep[e.g.][]{Filiz13}.

We also find that there is a strong correlation between the \civ
BAL-trough EW and the \siiv BAL-trough EW. The Spearman correlation
coefficient is 0.7 ($P_{null}=0.003$) for $EW_{1}$ and 0.8
($P_{null}=7\times 10^{-6}$) for $EW_{2}$.

\subsection{The continuum versus the spectral index $\alpha$}

\begin{figure}
\begin{center}
\includegraphics[height=8cm,angle=-90]{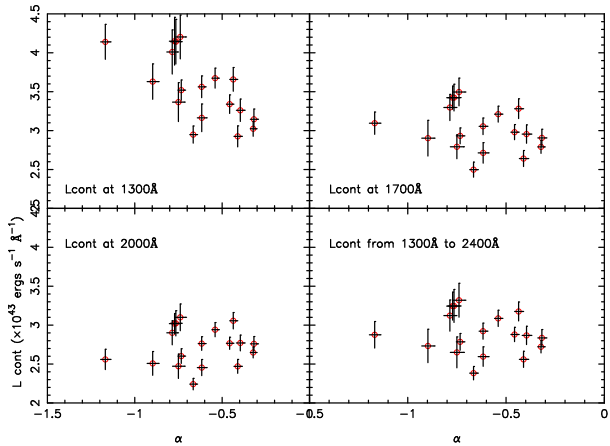}
\caption{The continuum versus the spectral index $\alpha$. The
continuum is measured at 1300 \AA\ (left top), at 1700 \AA\ (right
top), at 2000 \AA\ (left bottom),and from 1300 \AA\ to 2400 \AA\ (right
bottom). The correlation becomes stronger when the continuum
luminosity is measured for the bluer bands.}
\end{center}
\end{figure}

For the slope variation in QSOs, some investigations found that spectra of QSOs with
low redshift are bluer during their brighter phases \citep[e.g.][]{Vanden01, Pu06}. However, with two-epoch
variation, it was recently found that the spectra of half of the
QSOs appear redder during their brighter phases, especially for
high-z QSOs \citep{Bian12}. From Fig. 1, we find that the
variability is larger for the blue side in the rms or rms/mean
spectra. The slope $\alpha$ changes about 260\% (from -0.32 to
-1.16, Table 1). In Fig. 6, we show the relation between the
continuum and $\alpha$. In order to avoid accidental conclusions, we
measured four different continuum luminosities from the mean
power-law luminosity at different monochromatic wavelengths or
wavelength interval: (1) 1300 \AA\ to 2400 \AA; (2) 1300 \AA; (3)
1700 \AA; (4) 2000 \AA. The continuum luminosity (from 1300 \AA\ to
2400 \AA) changed by about 38\%. The Spearman correlation coefficients
between the continuum luminosity and $\alpha$ are -0.69 (0.001),
-0.36 (0.14), -0.001 (0.997), -0.30 (0.23) (from top to bottom and
from left to right in Fig. 6), where the values in brackets are
the probabilities of the null hypothesis. The correlation becomes stronger
when the continuum luminosity is measured for the bluer bands. This
correlation shows that this QSO has a steeper UV slope when
its luminosity increases.  This kind of negative correlation is
consistent with some previous studies for PG QSOs \citep[e.g.][]{Pu06}.

\subsection{EW versus the spectral index $\alpha$}

\begin{figure}
\begin{center}
\includegraphics[height=8cm,angle=-90]{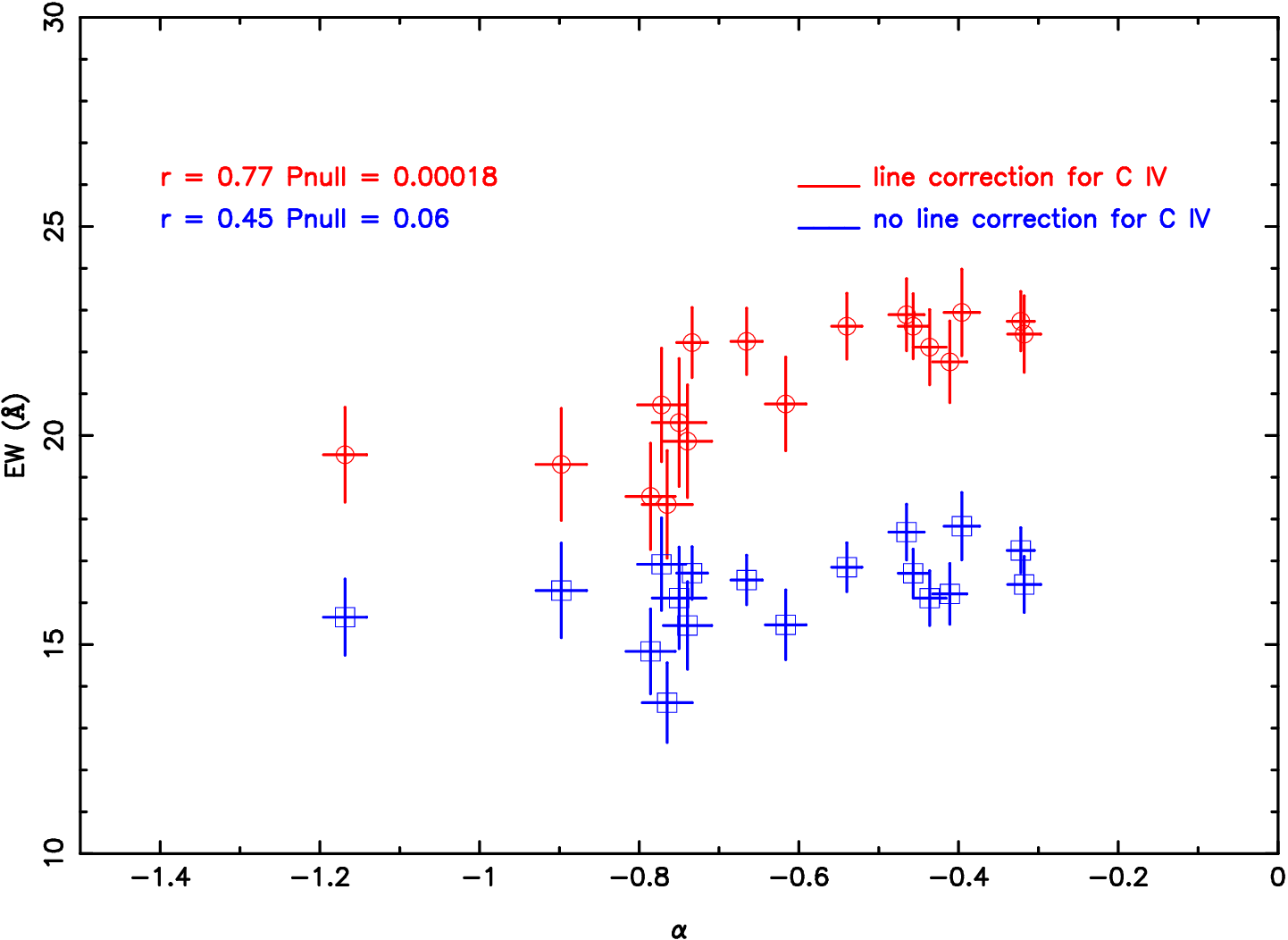}
\includegraphics[height=8cm,angle=-90]{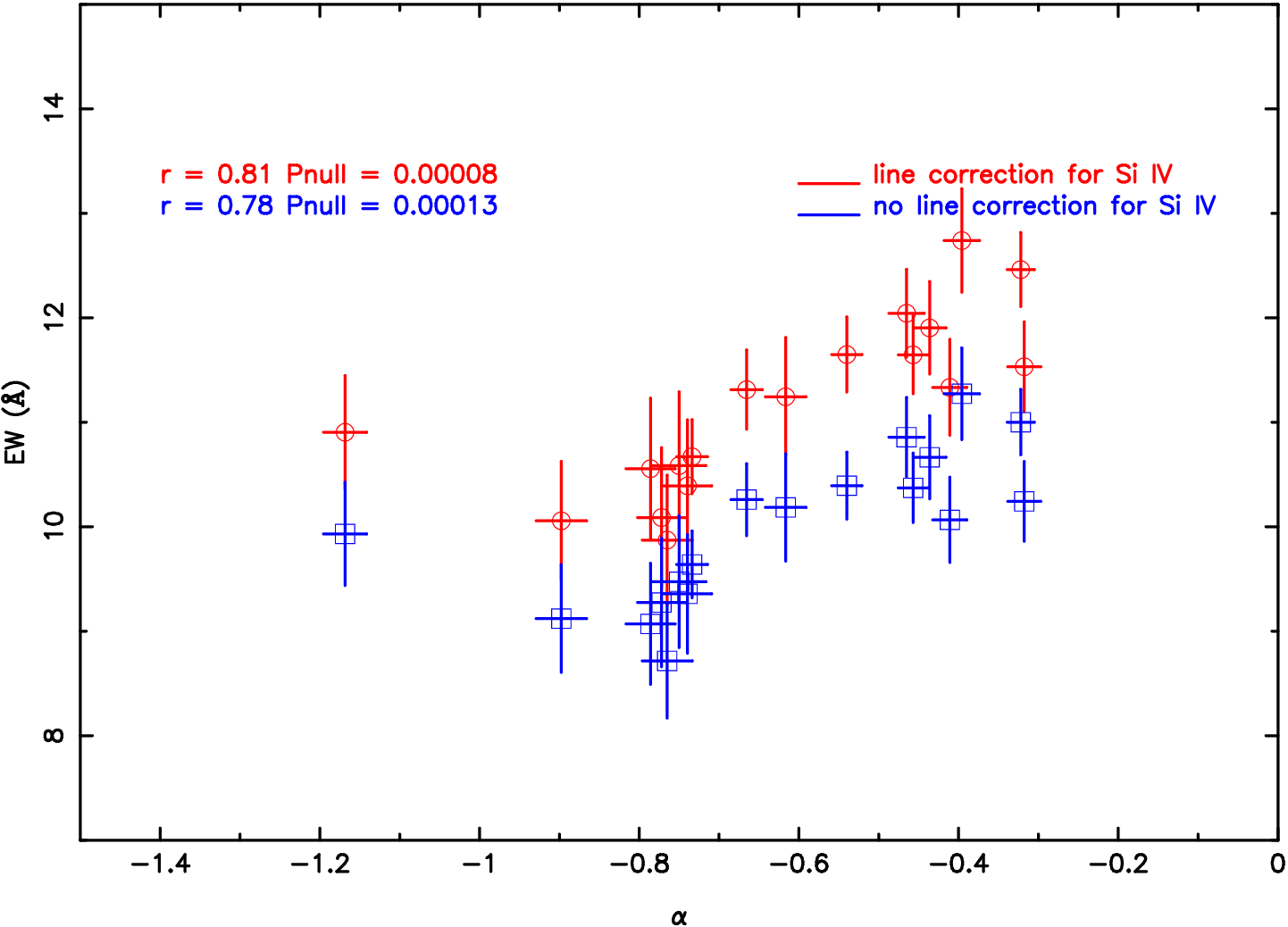}
\caption{ The BAL-trough EW versus spectral index $\alpha$ for the \civ BAL (top), and
for the \siiv BAL (bottom). The blue square is for the case based on the
 power-law continuum fit. The red circle is for the case based
on the pseudo-continuum fit of the power-law plus the emission lines.
}
\end{center}
\end{figure}

Fig. 7 displays the relation between the BAL-trough EW and the spectral
index $\alpha$ for \civ (top) and \siiv (bottom). The blue squares
denote the $EW_{1}$ data, i.e., based on the power-law continuum.
Red circles denote the $EW_{2}$ data, i.e., based on the
pseudo-continuum of the power-law plus the emission lines. Due to the
correction of the emission lines, the value of $EW_{2}$ is larger
than $EW_{1}$. The correlation between the EW and $\alpha$ is
strong. The
correlation coefficient is 0.45 (0.06) for \civ $EW_{1}$, 0.77
(0.00018) for \civ $EW_{2}$, 0.78 (0.00013) for \siiv $EW_{1}$, and
0.81 (0.00008) for \siiv $EW_{2}$, where the values in brackets are
the probabiliies of the null hypothesis. With the correction
from the emission lines creating deeper troughs, stronger correlations are found for the
\civ and \siiv BAL-troughs. We list these correlation coefficients in
Table 2.

This result shows that the BAL-trough EW becomes larger when the UV
slope becomes flatter, implying that the outflow has the effect of
reddening the observed spectrum. Recently, \cite{Baskin13} used
the mean SDSS spectra constructed by selecting on various QSO
properties to investigate the outflow dependence on those
properties. They found that the BAL-trough becomes deeper for QSOs
with low \heii $\lambda 1640$ \AA\ EW or flatter UV slope. Our
result is consistent with theirs. The intrinsic dust
Small-Magellanic-Cloud-like (SMC-like) reddening in QSO spectra is
discussed in some studies \citep[e.g.][]{Baskin13}. For this
single BAL QSO, the larger BAL-trough EW with flatter spectrum
implies an important role for the ionizing continuum in the trough
variation.

The origin of BALs is commonly thought to be the result of subtending a
 disk wind in the line of sight. The disk wind may be subject to
different physical effects in
different zones of its outflow velocity structure. It was found that the
high-velocity components of the \civ BAL-trough change much more than the
low-velocity part with the change of \heii 1640 \AA\ EW, which is
a measurement of the extreme UV continuum (Baskin et al. 2013).
Here we divide the total trough EW into two parts, for high-velocity and
low-velocity. For the \civ BAL-trough, the wavelength range is
between 1450 \AA\ and 1500 \AA\ ($\sim$ 20000 \kms - 10000 \kms) for
the high-velocity part, and between 1500 \AA\ and 1549 \AA\ ($\sim$
10000 \kms - 0 \kms) for the low-velocity part. For the \siiv
BAL-trough, the wavelength range is between 1340 \AA\ and 1355 \AA\
($\sim$ 13000 \kms - 9700 \kms) for the high-velocity part, and between
1355 \AA\ and 1400 \AA\ ($\sim$ 9700 \kms - 0 \kms ) for the
low-velocity part.

For the \civ BAL, the Spearman coefficients for the correlation
between the divided BAL-trough EWs and the spectral index $\alpha$
are 0.64 (0.0042) for the high velocity part and 0.87 (0.000003) for
the low-velocity part. The Spearman correlation coefficients are
0.75 (0.0003) for the high-velocity part of the \siiv BAL and 0.82
(0.00003) for the low-velocity part, where the values in brackets
are the probabilities of the null hypothesis. It seems that the
correlation between the spectral index and EW for the low-velocity
part is slightly stronger than that for the high-velocity part. That
result is consistent with the study of \citep{Zhang14}, where
they found that the UV slope affects the low-velocity part of
outflows much more.

\subsection{EW versus the continuum}

\begin{figure}
\begin{center}
\includegraphics[height=8cm,angle=-90]{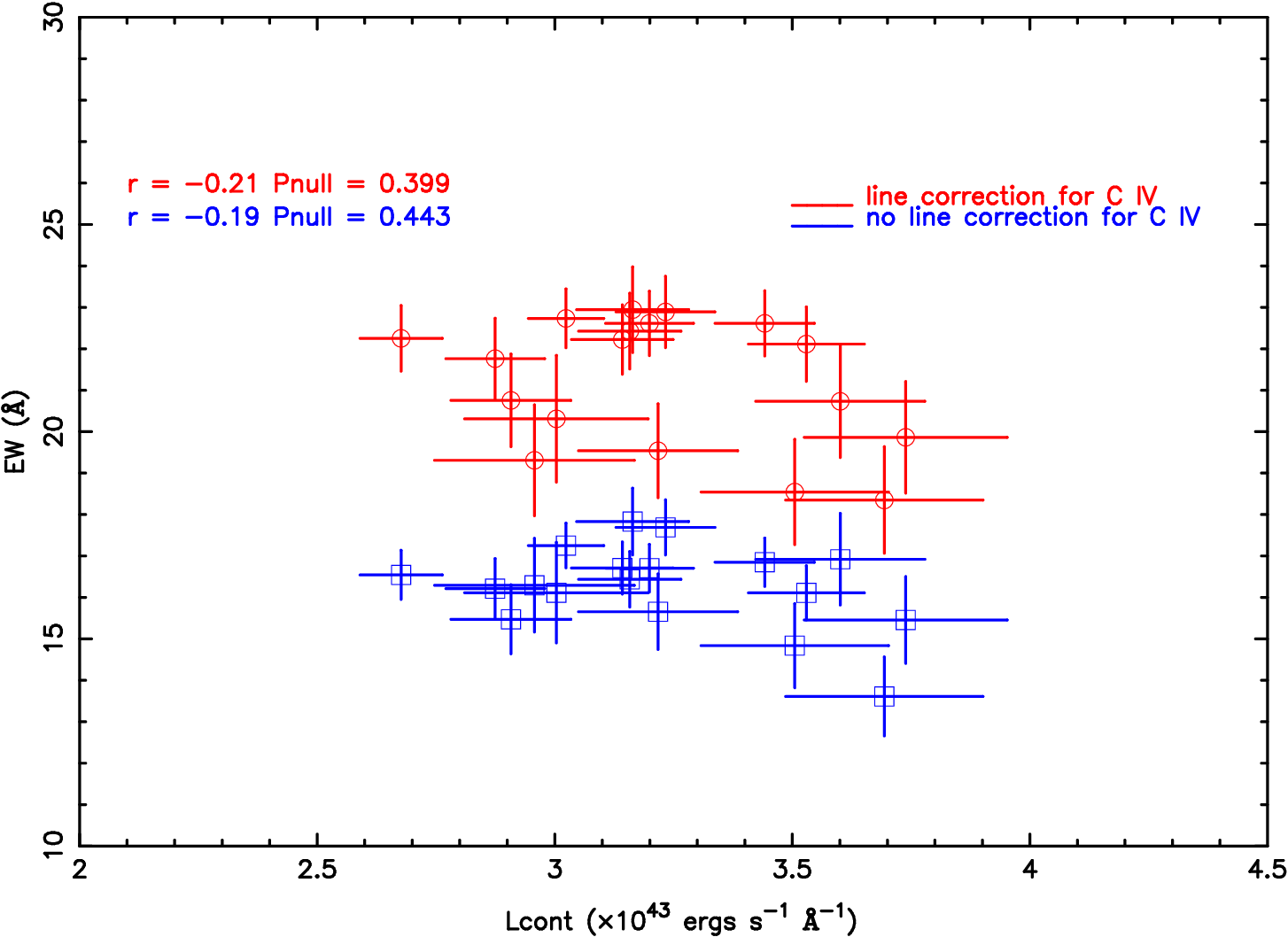}
\includegraphics[height=8cm,angle=-90]{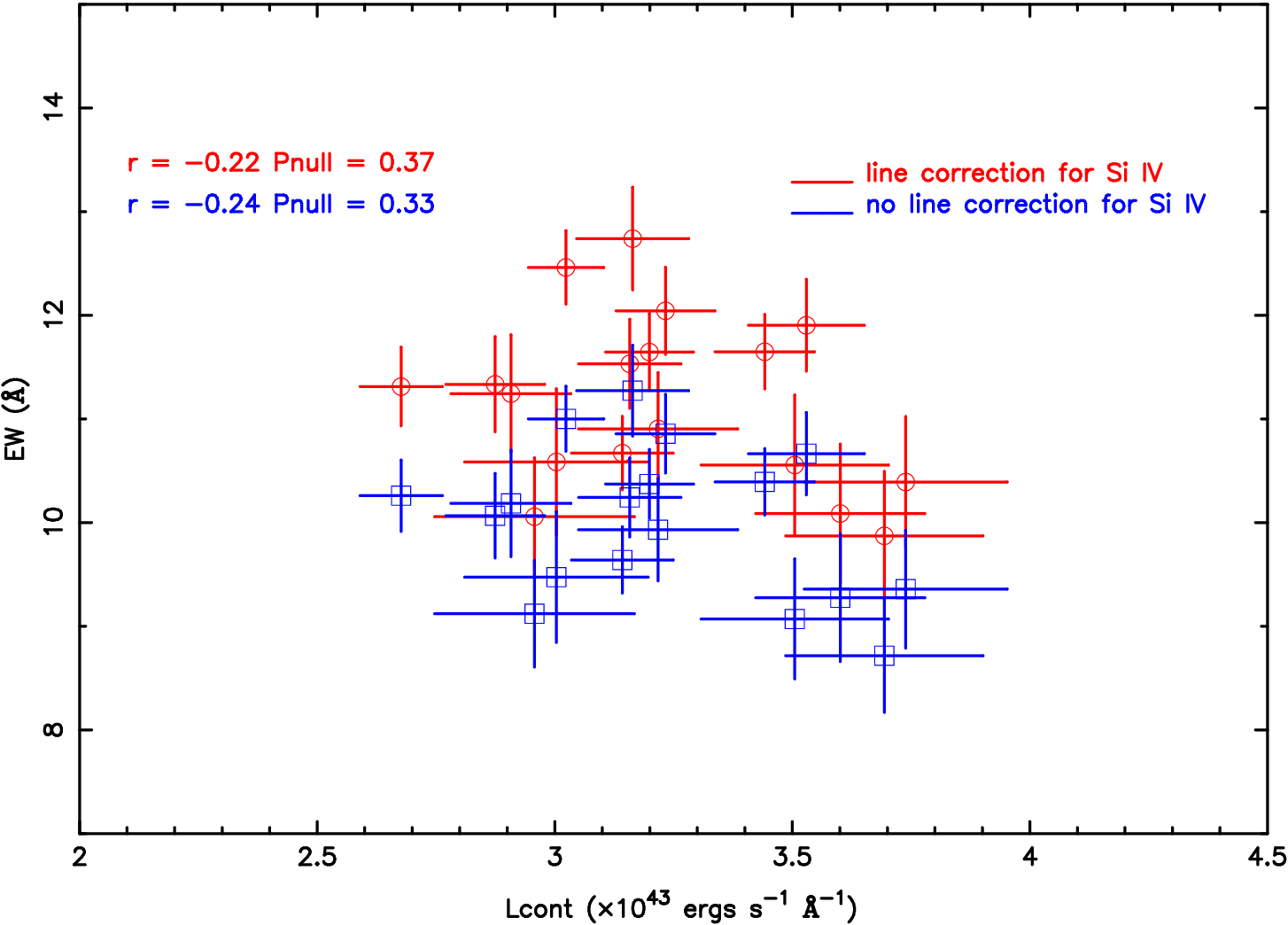}
\caption{ The EW versus the continuum luminosity for the \civ BAL (top),
and for the \siiv BAL (bottom). The symbols are the same as in Fig. 7.}
\end{center}
\end{figure}

\begin{figure}
\begin{center}
\includegraphics[height=7cm,angle=-90]{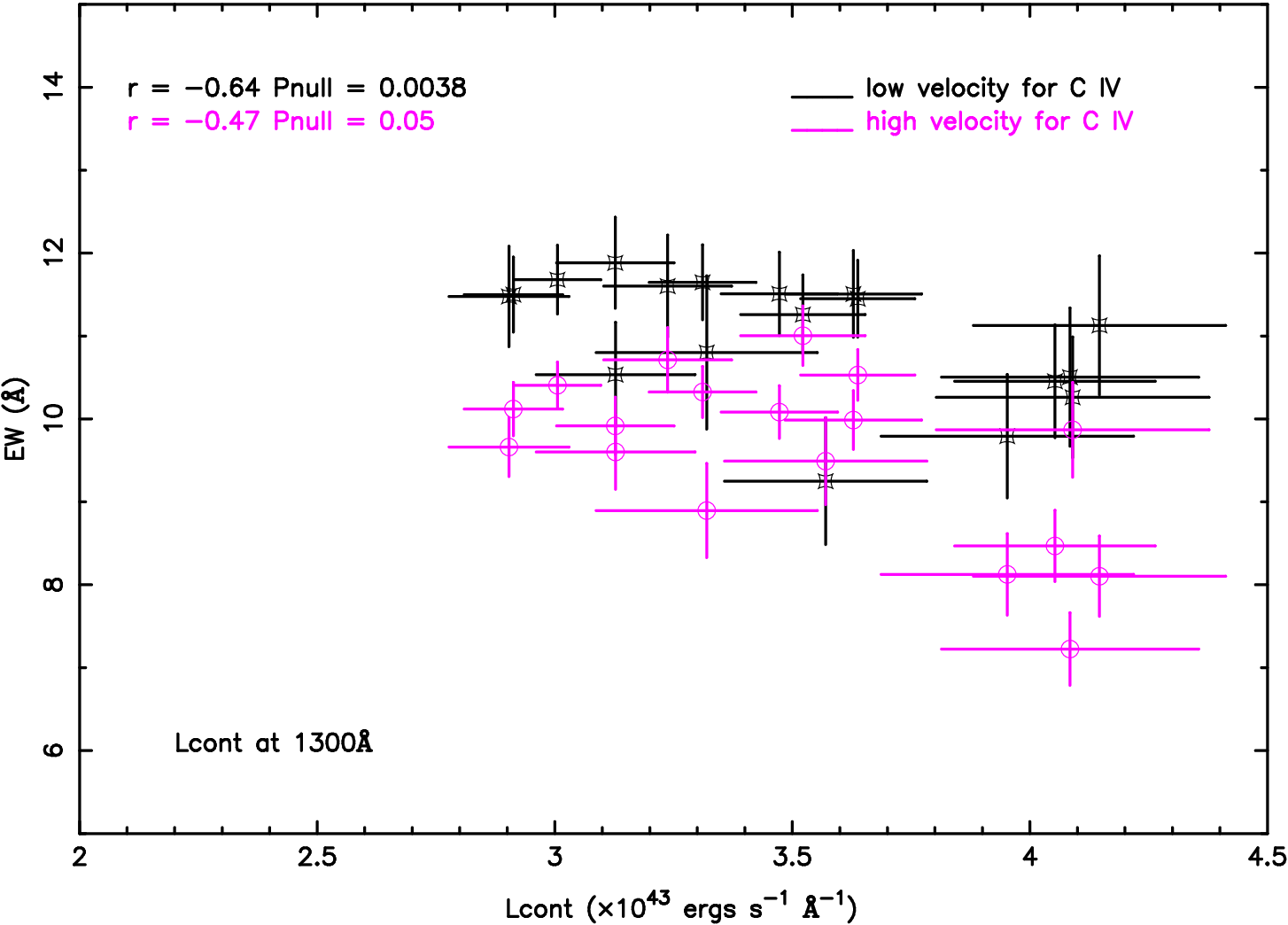}
\includegraphics[height=7cm,angle=-90]{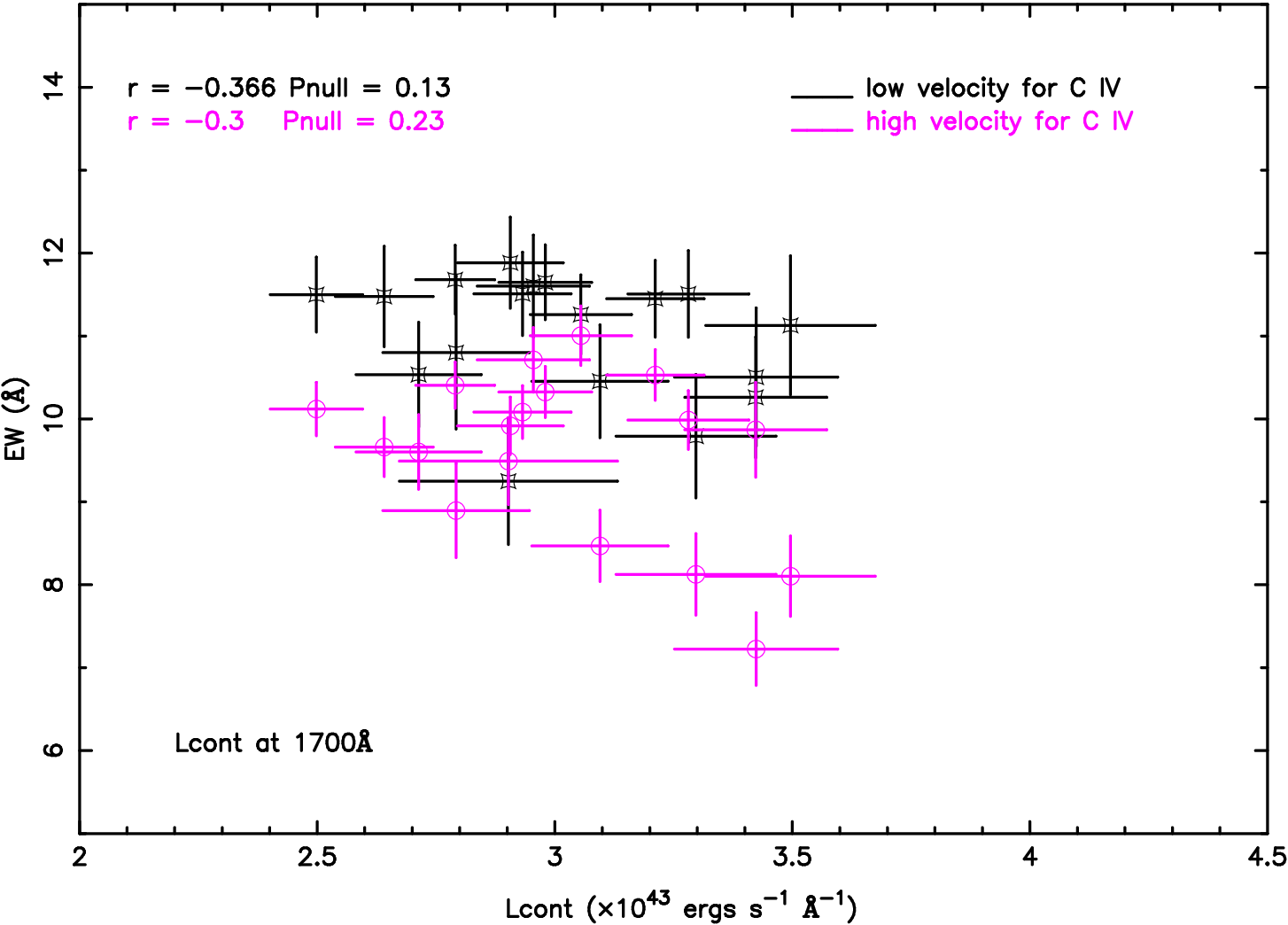}
\includegraphics[height=7cm,angle=-90]{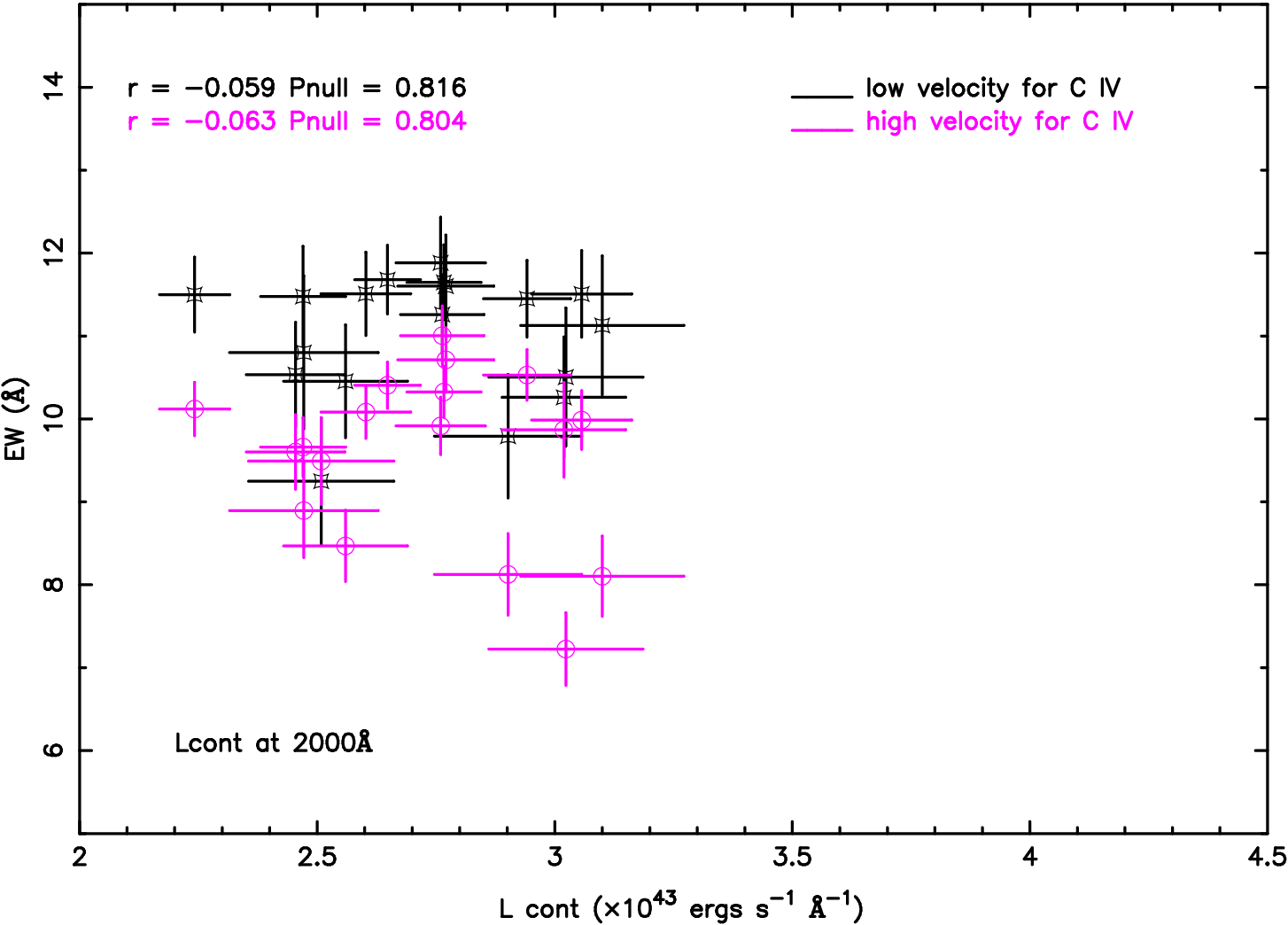}
\caption{\civ BAL EW vs the continuum, where the continuum luminosity is
measured at 1300 \AA\ (top), at 1700 \AA\ (middle) and 2000 \AA\
(bottom), respectively. The black star is for the low-velocity part
of the BAL-trough and the pink circle is for the high-velocity part of
the BAL-trough.}
\end{center}
\end{figure}

The dependence of the EW variability on QSO properties, such as
the continuum luminosity, is still a question for debate (e.g., Gibson 2008;
Filiz Ak et al. 2013; Baksin et al. 2013). Fig. 8 displays the
correlation between BAL-trough EW and the continuum luminosity for
\civ (top panel) and \siiv (bottom panel). As in Fig. 7, blue
squares denote the $EW_{1}$ data and red circles denote the $EW_{2}$
data. The correlation between the EW and the continuum is weak. The
Spearman coefficient is -0.19 (0.443) for \civ $EW_{1}$, -0.21
(0.399) for \civ $EW_{2}$, -0.24 (0.33) for \siiv $EW_{1}$, and
-0.22 (0.37) for $EW_{2}$, where the values in brackets are the
probabilities of the null hypothesis. The correlation between EW and the
continuum is weak with a large probability of the null hypothesis. There
is no significant correlation found in this QSO. It is consistent
with the study of \cite{Filiz13} who found no
significant evidence for EW variability of \civ BAL-trough driven by
the QSOs bolometric luminosity. We also calculated the Spearman
coefficients between EW and the continuum for the high-velocity and
low-velocity parts separately. The correlation coefficient between EW
and the continuum is about -0.22 (0.39) (high-velocity) and -0.28
(0.26) (low-velocity) for the \civ BAL-trough, and  is -0.35 (0.15)
(high-velocity) and -0.22 (0.39) (low-velocity) for the \siiv
BAL-trough, where the values in brackets are the probabilities of the
null hypothesis. There is still no significant correlation between
EW and the continuum for either the high-velocity or low-velocity
parts.

We also measured the continuum luminosity at 1300 \AA, 1700 \AA,
2000 \AA, respectively (Fig. 9). The Spearman correlation
coefficient between EW and the continuum luminosity (1300 \AA) for the
\civ BAL is about -0.64 (0.0038) (low-velocity) and -0.47 (0.05)
(high-velocity). For the continuum at 1700 \AA, the correlation
coefficient is -0.366 (0.13) (low-velocity) and -0.3 (0.23)
(high-velocity). For the continuum at 2000 \AA, the correlation
coefficient is -0.059 (0.816) (low -velocity) and -0.063
(0.804) (high-velocity). The values in brackets are the
probabilities of the null hypothesis. We find that the correlation
becomes stronger when the continuum luminosity is measured in the bluer
bands. The measured continuum luminosity in the bluer bands is more affected by
dust extinction. This result is consistent with the the correlation between
the BAL-trough EW and the spectral index.

\subsection{The maximum velocity of outflow}
\begin{figure}
\begin{center}
\includegraphics[width=6cm,angle=-90]{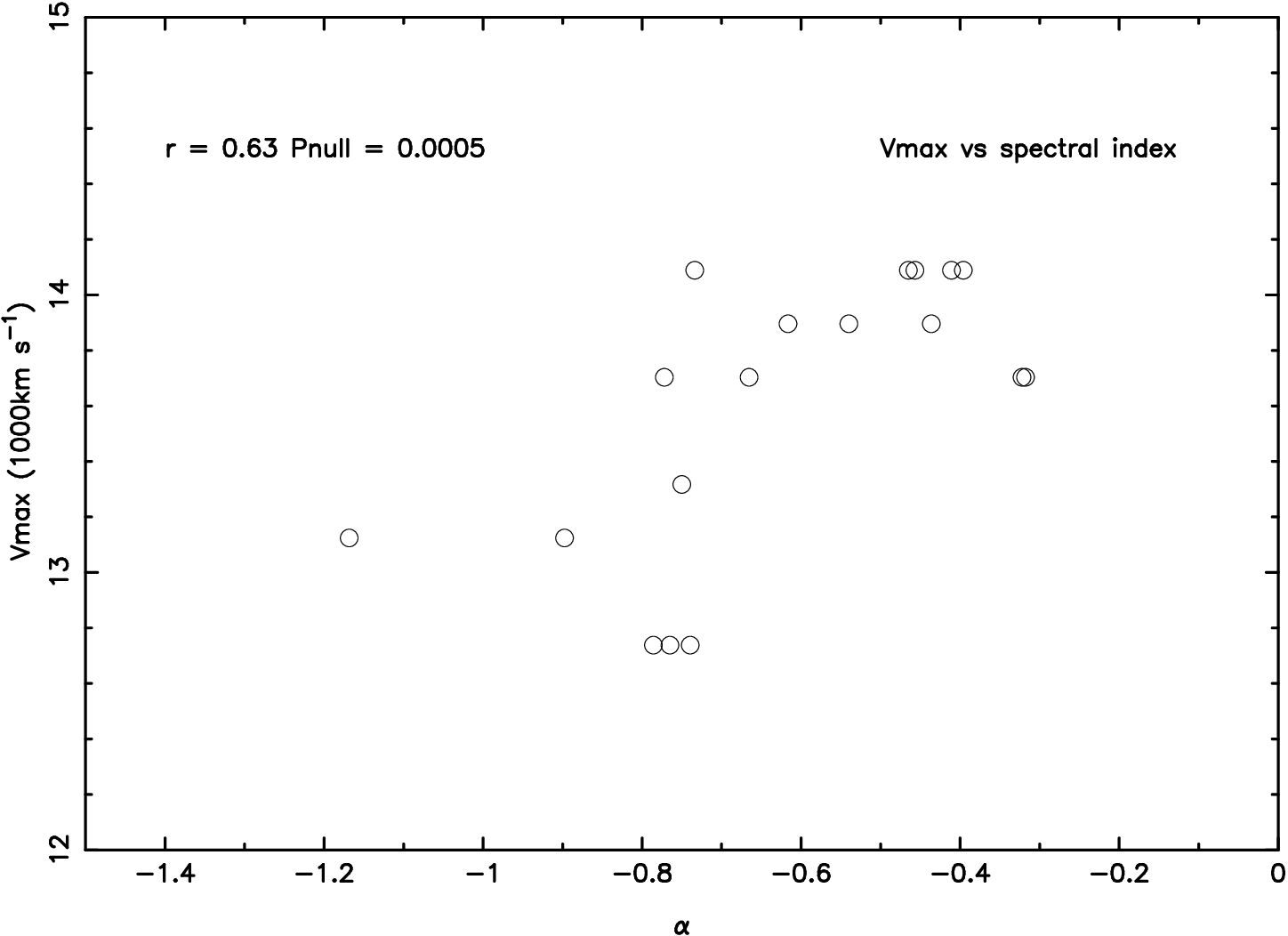}
\includegraphics[width=6cm,angle=-90]{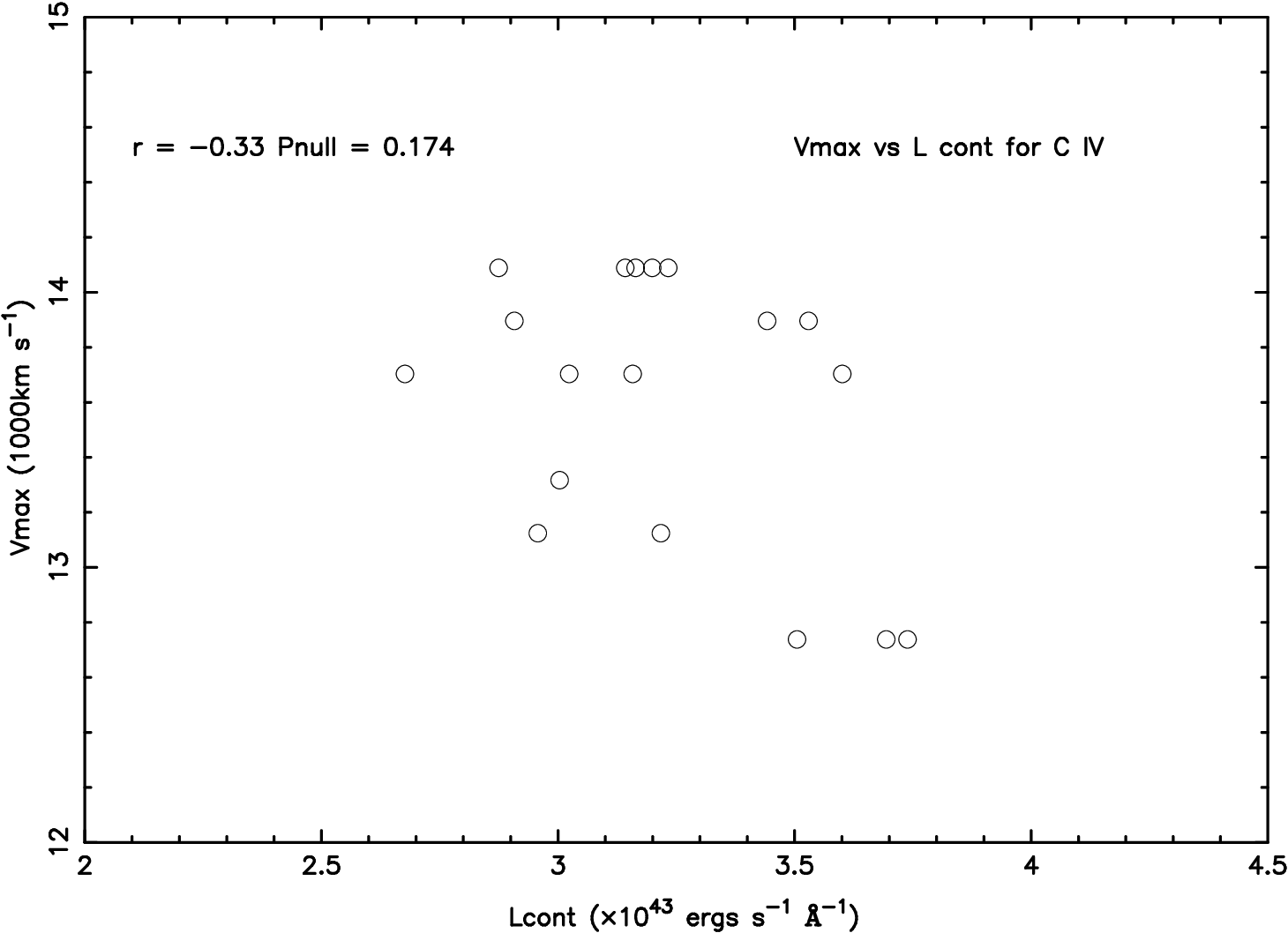}
\caption{The maximum velocity of outflow versus spectral index $\alpha$ and the continuum
(measured from  1300\AA\ to 2400\AA ). The Spearman correlation coefficient for
the maximum velocity and spectral index $\alpha$ is 0.63
with $P_{null}=0.0005$. The Spearman correlation coefficient for
the maximum velocity and the continuum is -0.33
with $P_{null}=0.174$.}
\end{center}
\end{figure}

The \civ maximum velocity of outflow $V_{max}$ for the 18
observations changed by about 10\% (Table 2). In Fig. 10, we show the
relation between  $V_{max}$ and the spectral index $\alpha$. There
is a moderately strong correlation between them with Spearman
coefficient of 0.63 ($P_{null}=0.0005$). This moderately strong
correlation suggests larger extinction for larger maximum velocity
of outflow. As we find a strong correlation between the BAL-trough
EW and the spectral index, there should be a correlation between $V_{max}$ and
the \civ BAL-trough EW, and indeed, the Spearman coefficient is 0.78
($P_{null}=0.00013$). For flatter UV slope and the correspondingly larger
BAL-tough, $V_{max}$ becomes larger. For the correlation between
 $V_{max}$ and the continuum(measured from  1300\AA\ to 2400\AA ), the Spearman coefficient is -0.33
($P_{null}=0.174$).  A weak negative correlation seems to exist
between $V_{max}$ and the continuum with a probability of 82.6\%. The
weak negative correlation between $V_{max}$ and the luminosity for this
one QSO is not consistent with the positive correlation found by others \citep[e.g.][]{Laor02}.
However, we should note that our correlation between
 $V_{max}$ and the luminosity is weak. More data are needed in
the future for this study.

\section{SUMMARY}
The variability of the broad absorption lines is investigated for a
broad absorption line QSO, SDSS J022844.09+000217.0 (z = 2.719),
with 18 SDSS/BOSS spectra covering 4128 days in the observational
frame. The main conclusions can be summarized as follows:

(1) The light curves for the BAL \civ EW, the continuum, and the
spectral index are given. The EW for the \civ BAL-trough changes by about
25\% (from 18.1 \AA\ to 22.9 \AA). The $\alpha$ changes by about
260\% (from -0.32 to -1.16). The continuum luminosity changes by
about 38\% (from 2.67 to 3.69, in units of $10^{43} \ergs
\AA^{-1}$).

(2) Through the rms-to-mean ratio spectrum, the relative flux change
increases as the frequency increases, from about 10\% on the red side,
to about 15\% on the blue side. The fractional variability is larger for
the absorption trough than for the continuum and the emission lines.

(3) There is a strong positive correlation between the \civ
BAL-trough EW and UV spectral index $\alpha$, as well as for the
\siiv BAL-trough. The larger BAL-trough EW with flatter spectrum
implies that the outflow has the effect of reddening the observed
spectrum.  The Spearman coefficient between the spectral index and
BAL-trough EW for the low-velocity part is slightly larger than that
for the high-velocity part. The strong correlation between the
BAL-trough EW and the spectral index for this one QSO suggests that
dust is intrinsic to outflows.

(4)There is no significant correlation between the \civ BAL-trough
EW and the continuum in this QSO, nor for the \siiv BAL-trough. We
do not find significant evidence for EW variability of the \civ
BAL-trough driven by the QSO's bolometric luminosity. We find that
the correlation becomes stronger when the continuum luminosity is
measured in the bluer bands. The weak correlation between the BAL
variability and the continuum luminosity for this one QSO implies
that the BAL-trough variation is not dominated by photoionization.

(5)There is a moderately strong correlation between the maximum
velocity of the \civ BAL-trough and the spectral index $\alpha$ for
this QSO. However, there is no significant correlation between the
maximum velocity of outflow and the continuum. More observations are
needed in the future for this study.

\section{ACKNOWLEDGMENTS}
We are very grateful to the anonymous referee for her/his
instructive comments which significantly improved the content of the
paper. We thank Richard Green F. very much for the
grammar-correction and comments in the paper. This work has been
supported by the National Science Foundations of China (No.
11373024; 11173016; 11233003).

\end{document}